\documentclass[12pt]{article}

\global\arraycolsep=1pt
\usepackage{amsmath,amssymb}

\newcommand{\beq}{\begin{equation}}
\newcommand{\eeq}{\end{equation}}
\newcommand{\beqa}{\begin{eqnarray}}
\newcommand{\eeqa}{\end{eqnarray}}

\renewcommand{\thefootnote}{\fnsymbol{footnote}}

\begin{document}

\begin{flushright}
May, 2008 \\
OCU-PHYS 297 \\
\end{flushright}
\vspace{5mm}

\begin{center}
{\bf\Large
Generalized Kerr-NUT-de Sitter metrics \\in all dimensions
}
\end{center}

\begin{center}

\vspace{10mm}

Tsuyoshi Houri$^a$\footnote{
\texttt{houri@sci.osaka-cu.ac.jp}
}, 
Takeshi Oota$^b$\footnote{
\texttt{toota@sci.osaka-cu.ac.jp}
} and
Yukinori Yasui$^a$\footnote{
\texttt{yasui@sci.osaka-cu.ac.jp}
}

\vspace{10mm}

\textit{
${}^a$Department of Mathematics and Physics, Graduate School of Science,\\
Osaka City University\\
3-3-138 Sugimoto, Sumiyoshi,
Osaka 558-8585, JAPAN
}
\vspace{5mm}

\textit{
${{}^b}$Osaka City University
Advanced Mathematical Institute (OCAMI)\\
3-3-138 Sugimoto, Sumiyoshi,
Osaka 558-8585, JAPAN
}

\vspace{5mm}

\end{center}
\vspace{8mm}

\begin{abstract}
We classify all spacetimes with a closed rank-2 conformal Killing-Yano tensor.
They give a generalization of Kerr-NUT-de Sitter spacetimes.
The Einstein condition is explicitly solved and written as an indefinite integral.
It is characterized by a polynomial in the integrand.
We briefly discuss the smoothness conditions 
of the Einstein metrics over compact Riemannian manifolds.
\end{abstract}

\vspace{25mm}

\newpage

\renewcommand{\thefootnote}{\arabic{footnote}}
\setcounter{footnote}{0}


The $D$-dimensional Kerr-NUT-de Sitter metric was 
constructed by Chen-L\"u-Pope \cite{CLP}.
The metric is the most general known solution 
describing the higher-dimensional rotating black hole spacetime 
with NUT parameters. 
It takes the form
\begin{equation}
g= \sum_{\mu=1}^{n} \frac{d x_{\mu}^2}{Q_\mu(x)}+
\sum_{\mu=1}^{n} Q_{\mu}(x) \left( \sum_{k=0}^{n-1} \sigma_{k}(\hat{x}_{\mu})
d \psi_k \right)^2+\frac{\varepsilon c}{\sigma_n}\left( \sum_{k=0}^{n} \sigma_k
d \psi_k \right)^2, 
\end{equation} 
where $D=2n+\varepsilon$ ($\varepsilon=0$ or $1$).
The functions $Q_{\mu}~(\mu=1,2, \cdots , n)$ are given by
\begin{equation}
Q_{\mu}(x)=\frac{X_{\mu}}{U_{\mu}},~~~
U_{\mu}=\prod_{\stackrel{\scriptstyle \nu=1}{(\nu \ne \mu)}}^{n}(x_{\mu}^2-x_{\nu}^2), 
\end{equation}
where $X_{\mu}=X_{\mu}(x_{\mu})$ is an arbitrary function depending 
on one coordinate $x_{\mu}$.
The $\sigma_k$ and $\sigma_k(\hat{x}_{\mu})$ are the $k$-th elementary symmetric functions
of $\{ x_{1}^2, \cdots , x_n^2\}$ and $\{x_{\nu}^2 : \nu \ne \mu \}$ respectively: 
\beq
\prod_{\nu=1}^{n}(t-x_{\nu}^2)=\sigma_0t^n-\sigma_1 t^{n-1}+ \cdots + (-1)^n \sigma_n,
\eeq
\beq
\prod_{\stackrel{\scriptstyle \nu=1}{(\nu \ne \mu)}}^{n}
(t-x_{\nu}^2)=\sigma_0(\hat{x}_{\mu}) t^{n-1}-\sigma_1(\hat{x}_{\mu}) t^{n-2}
+ \cdots + (-1)^{n-1} \sigma_{n-1}(\hat{x}_{\mu}).
\eeq

The metric satisfies the Einstein equation $Ric(g)= \Lambda g$ 
if and only if $X_{\mu}$ takes the form \cite{CLP,HHOY},
\begin{equation}
(a)~~ \varepsilon=0:
X_{\mu}=\sum_{k=0}^{n} c_{k} x_{\mu}^{2k}+b_{\mu} x_{\mu},~~~
(b)~~ \varepsilon=1:
X_{\mu}=\sum_{k=0}^{n} c_{k} x_{\mu}^{2k}+b_{\mu}
+\frac{(-1)^n c}{x_{\mu}^2},
\end{equation}
where $c, c_k$ and $b_{\mu}$ are free parameters. 
This class of metrics gives the Kerr-NUT-de Sitter metric \cite{CLP},
and the solutions in \cite{MP,HHT,GLPP1,GLPP2,CLP1} are 
recovered by choosing special parameters.

It has been shown in \cite{FK,KF} that the Kerr-NUT-de Sitter spacetime 
has a rank-2 closed conformal Killing-Yano~(CKY) tensor.
This tensor generates the tower of Killing-Yano and Killing tensors, 
which implies complete integrability of geodesic equations \cite{PKVK}
and complete separation of variables for the Hamilton-Jacobi, 
Klein-Gordon \cite{FK,FKK} and Dirac equations \cite{OY2}.
Various aspects related to the integrability have been extensively studied
in \cite{KKPF,KKPV,HOY,HOY3,KF07,SK,CFK,KFK08}. For reviews on these subjects,
see, for example, \cite{flor08,FK08}.

This property leads to a natural question whether there are other geometries with  
such a CKY tensor. The following result was proved in \cite{HOY3}.\\

\noindent
{\bf{Theorem 1.}} Let us assume the existence of a 
non-degenerate rank-$2$ CKY tensor $h$ for $D$-dimensional
spacetime $(M ,g)$ satisfying the conditions\footnote{
Recently, it was proved that the assumptions of $(a2)$ and $(a3)$ 
are superfluous because they
follow from the existence of the closed CKY tensor
\cite{KFK08}.},
\[ (a1)~ dh=0,~~(a2)~{\cal{L}}_{\widehat{\xi}}g=0,~~(a3)~{\cal{L}}_{\widehat{\xi}}h=0 . \]
Then, $M$ is only the Kerr-NUT-de Sitter spacetime.\\

The rank-2 CKY tensor $h=(h_{ab})$ is a 
2-form defined by the equation \cite{tac}
\begin{equation}
\nabla_a h_{bc}+\nabla_b h_{ac}=2 \widehat{\xi}_c g_{ab}
-\widehat{\xi}_a g_{bc}-\widehat{\xi}_b g_{ac},
\end{equation}
where the associated vector $\widehat{\xi}=\widehat{\xi}^a\partial_a$
of $h$ is given by $
\widehat{\xi}_a=(1/(D-1)) \nabla^b h_{ba}.
$
By introducing the following orthonormal frame 
\begin{equation}
e^{\mu}=\frac{dx_{\mu}}{\sqrt{Q_{\mu}}},
~~e^{n+\mu}=\sqrt{Q_{\mu}}\sum_{k=0}^{n-1} \sigma_{k}(\hat{x}_{\mu})d \psi_k ,~~
\varepsilon \, e^{2n+1}=\varepsilon\, \sqrt{\frac{c}{\sigma_n}}\sum_{k=0}^{n} \sigma_k
d \psi_k 
\end{equation}
for the metric (1), the CKY tensor can be written as
\begin{equation}
h=\sum_{\mu=1}^n x_{\mu}e^{\mu} \wedge e^{n+\mu}
= d\left( \frac{1}{2} \sum_{k=0}^{n-1} \sigma_{k+1} d \psi_k \right).
\end{equation}
In \cite{HOY3} we required that the eigenvalues $x_{\mu}$ of $h$ are functionally independent in some
spacetime domain,~i.e. $x_{\mu}$ are non-constant independent functions.
In this paper we do not assume the functional
independence of the eigenvalues, and hence 
the CKY tensor generally has the
non-constant eigenvalues and the constant ones. The metric may be locally
given as a Kaluza-Klein metric on the bundle over K\"ahler manifolds whose
fibers are Kerr-NUT-de Sitter spacetimes.


Let $(M, g)$ be a $D$-dimensional spacetime 
with a closed rank-$2$ CKY tensor $h$.
Let $x_{\mu}$ ($\mu=1,\cdots, n$) 
and 
$\xi_i$ 
($i=1,\cdots, N$) be
the non-constant eigenvalues and the non-zero constant ones of $h$, respectively.
Suppose the eigenvalues of the ``square of the CKY tensor" $Q=(Q^a{}_b) = ( - h^{a}{}_c h^c{}_b )$ 
have the following multiplicities:
\begin{equation}
\{
\underbrace{x_1^2, \dotsm,  x_1^2}_{2\ell_1},
\dotsm, 
\underbrace{x_n^2, \dotsm, x_n^2}_{2\ell_n}, 
\underbrace{\xi_1^2, \dotsm, \xi_1^2}_{2m_1},
\dotsm,
\underbrace{\xi_N^2, \dotsm, \xi_N^2}_{2m_N}, 
\underbrace{0,\dotsc, 0}_{K} \},
\end{equation}
where $D=2 (|\ell| + |m| )+ K$.
Here $|\ell| = \sum_{\mu=1}^n \ell_{\mu}$ and $|m|=\sum_{i=1}^N m_i$. 

Analyses for the non-degenerate and some degenerate cases with $|m|=0$
can be found in \cite{CFK}.

We can  show the following results \cite{HOY2}:\\

\noindent
{\bf{Lemma.}} It must hold that $\ell_{\mu}=1$ for all $\mu=1,2,\dotsc, n$.\\

\noindent
{\bf{Theorem 2.}} The metric and the CKY tensor take the forms
\begin{eqnarray} \label{gCKY}
g &=& 
\sum_{\mu=1}^{n} \frac{d x_{\mu}^2}{P_\mu (x)}+
\sum_{\mu=1}^{n} P_{\mu}(x) \left( \sum_{k=0}^{n-1} \sigma_{k}(\hat{x}_{\mu})
\theta_k \right)^2
+ \sum_{i=1}^{N}\prod_{\mu=1}^{n}(x_{\mu}^2- \xi_i^2)g^{(i)}+
\sigma_n \, g^{(0)},\\ 
\label{cCKY}
h&=&\sum_{\mu=1}^n x_{\mu} dx_{\mu} \wedge 
\left(\sum_{k=0}^{n-1} \sigma_{k}(\hat{x}_{\mu})
\theta_k \right)
+\sum_{i=1}^N \xi_i \prod_{\mu=1}^{n}(x_{\mu}^2- \xi_i^2)\omega^{(i)}.
\end{eqnarray}
The metrics $g^{(i)}$ are 
K\"ahler metrics on $2 m_i$-dimensional K\"ahler manifolds $M^{(i)}$
and $\omega^{(i)}$ the corresponding K\"ahler forms. The metric $g^{(0)}$ is, in general, any
 metric on a $K$-dimensional manifold $M^{(0)}$. 
But if $K=1$, 
$g^{(0)}$ can take the
special form:
\begin{equation}\label{g0sp}
\sigma_n \, g^{(0)}_{\mathrm{special}} = \frac{c}{\sigma_n} \left( \sum_{k=0}^{n} \sigma_k
\theta_k \right)^2.
\end{equation}
The functions $P_{\mu}$ are defined by
\begin{equation}
P_{\mu}(x)=\frac{X_{\mu}}
{\displaystyle x_{\mu}^K \prod_{i=1}^{N}(x_{\mu}^2-\xi_i^2)^{m_i}\, U_{\mu}}, \qquad
U_{\mu}=\prod_{\stackrel{\scriptstyle \nu=1}{(\nu \ne \mu)}}^{n}
(x_{\mu}^2-x_{\nu}^2)
\end{equation}
with an arbitrary function $X_{\mu}=X_{\mu}(x_{\mu})$ depending on $x_{\mu}$.
The 1-forms $\theta_k$ satisfy
\begin{equation}
d\theta_k+ 2\sum_{i=1}^N (-1)^{n-k} \xi_i^{2n-2k-1} \omega^{(i)}=0, \qquad
k=0,1,\dotsc, n-1+\varepsilon,
\end{equation}
where $\varepsilon=0$ for the general type and $\varepsilon=1$
for the special type.\\

\noindent
{\bf{Remark 1.}} Locally, $\omega^{(i)} = dA^{(i)}$, and 
$\theta_k = d \psi_k - 2 \sum_{i=1}^N (-1)^{n-k} \xi_i^{2(n-k)-1} A^{(i)}$.
The closed CKY tensor \eqref{cCKY}
can be rewritten in a manifestly closed form:
\begin{equation}
h = d \left( \frac{1}{2} \sum_{k=0}^{n-1} \sigma_{k+1} d \psi_k
+ \sum_{i=1}^N \xi_i \prod_{\mu=1}^n ( x_{\mu}^2- \xi_i^2  ) A^{(i)} \right).
\end{equation}

\noindent
{\bf{Remark 2.}} By sending one of constant eigenvalues, say $\xi_N$, to zero,
the metric of general type smoothly goes to a metric of  general type :
$N \rightarrow N-1$, $K \rightarrow K+2m_N$, $g^{(0)} \rightarrow
g^{(0)} + g^{(N)}$, where $g^{(N)}$ is the K\"{a}hler metric.
For special type, if we set $c=\xi_N^2$, $\psi_n=\varphi/\xi_N$,
and then take the limit\footnote{
The limit is different from the BPS limit \cite{CLPP,CLPP2}.
In this limit, the Sasakian manifold appears as a subspace of base space.
While in the BPS limit, the odd dimensional Kerr-NUT-de Sitter space (fiber space) goes
to a Sasakian manifold. This BPS limit was first done in \cite{CLPP,CLPP2}.}, 
$\xi_N \rightarrow 0$,  it goes to a metric of general type :
$N \rightarrow N-1$, $K=1 \rightarrow 2m_N+1$, $g^{(0)}$ part is given by
a Sasakian manifold, i.e., 
an $S^1$-bundle over the K\"{a}hler manifold,
\begin{equation} \label{Sasaki}
g^{(0)}_{\mathrm{special}} \rightarrow g^{(N)} + 
\left( d \varphi - 2 A^{(N)} \right)^2.
\end{equation}

\vspace{3mm}


In the following part, we consider the Einstein condition of the metric \eqref{gCKY}.
We introduce an orthonormal frame
$\{e^{A} \}=\{e^{\mu}, e^{n+\mu}, 
e^{\alpha}_{(i)}, e^{m_i+\alpha}_{(i)}, e^{\alpha}_{(0)} \}$ on $M$:
\begin{eqnarray}
e^{\mu}&=& \frac{dx_{\mu}}{\sqrt{P_{\mu}}},
~~e^{n+\mu}=\sqrt{P_{\mu}}\sum_{k=0}^{n-1} \sigma_{k}(\hat{x}_{\mu})\theta_k ,~~
e^{\alpha}_{(0)}= 
\sqrt{\sigma_n} \hat{e}^{\alpha}_{(0)}, \nonumber\\
\label{K0viel}
e^{\alpha}_{(i)}&=&
\left( \prod_{\mu=1}^{n} ( x_{\mu}^2- \xi_i^2 ) \right)^{1/2} \hat{e}^{\alpha}_{(i)},~~ 
e^{m_i+\alpha}_{(i)}=
\left( \prod_{\mu=1}^{n} ( x_{\mu}^2- \xi_i^2 ) \right)^{1/2} \hat{e}^{m_i+\alpha}_{(i)},~~ 
\end{eqnarray}
where 
$\{\hat{e}^{\alpha}_{(i)}, \hat{e}^{m_i+\alpha}_{(i)} \}_{\alpha=1,2,\dotsc, m_i}$ 
is an orthonormal frame of the
K\"ahler manifold $M^{(i)}$,
\begin{equation}
g^{(i)}=\sum_{\alpha=1}^{m_i}(\hat{e}^{\alpha}_{(i)}\otimes \hat{e}^{\alpha}_{(i)}+
\hat{e}^{m_i+\alpha}_{(i)} \otimes \hat{e}^{m_i+\alpha}_{(i)} ),~~
\omega^{(i)}=\sum_{\alpha=1}^{m_i} \hat{e}^{\alpha}_{(i)} \wedge
\hat{e}^{m_i+\alpha}_{(i)},
\end{equation}
and $\{ \hat{e}^{\alpha}_{(0)} \}_{\alpha=1,2,\dotsc,K}$ is an orthonormal frame of $M^{(0)}$.
For special type, we use 
\begin{equation}
e^{1}_{(0)} = \sqrt{\frac{c}{\sigma_n}} \sum_{k=0}^n \sigma_k \theta_k,
\end{equation}
instead of $ \{ e^{\alpha}_{(0)} \}$.

The CKY tensor \eqref{cCKY} is written as
\begin{equation}
h=\sum_{\mu=1}^n x_{\mu}e^{\mu} \wedge e^{n+\mu}
+ \sum_{i=1}^N \sum_{\alpha=1}^{m_i} \xi_i e^{\alpha}_{(i)} \wedge e^{m_i+\alpha}_{(i)}.
\end{equation}
It is convenient to introduce the following scalars
\begin{equation}
P_T^{[k]}(t):= || ( Q - t I)^{-k/2} \widehat{\xi} ||^2
=\widehat{\xi}_a \left( ( Q - t I)^{-k} \right)^a{}_b \widehat{\xi}^b
= \sum_{\mu=1}^n \frac{P_{\mu}}{(x_{\mu}^2 - t)^k}
+ \frac{ \varepsilon S}{(-t)^k},
\end{equation}
\begin{equation}
P_T=||\widehat{\xi}||^2 = \widehat{\xi}_a \widehat{\xi}^a= \sum_{\mu=1}^n P_{\mu}+ 
\varepsilon S
=P_T^{[0]}(t),
\qquad \varepsilon S=\varepsilon \frac{c}{\sigma_n}.
\end{equation}

The non-zero components of the Ricci tensor of the metric \eqref{gCKY} are calculated as
\begin{eqnarray}
\mathcal{R}_{\mu, \mu}
&=&\mathcal{R}_{n+\mu, n+\mu}\nonumber\\
&=&-\frac{1}{2} \frac{\partial^2 P_T}{\partial x_{\mu}^2}-\sum_{\rho \ne \mu}
\frac{1}{x_{\rho}^2-x_{\mu}^2} \left( x_{\rho} \frac{\partial P_T}{\partial x_{\rho}}
-x_{\mu} \frac{\partial P_T}{\partial x_{\mu}} \right)
-\frac{\varepsilon}{2 x_{\mu}} \frac{\partial P_T}{\partial x_{\mu}} \nonumber \\
&-&\sum_{i=1}^N m_i \left( x_{\mu} \frac{\partial}{\partial x_{\mu}} + 2 \right)
P_T^{[1]}(\xi_i^2), \nonumber\\
\mathcal{R}_{(\alpha,i),(\beta,i)}
&=&\mathcal{R}_{(m_i+\alpha,i),(m_i+\beta,i)} \nonumber \\
&=&\frac{\hat{\mathcal{R}}^{(i)}_{\alpha \beta}}{\prod_{\mu=1}^{n}(x_{\mu}^2- \xi_i^2)}
- \delta_{\alpha \beta} \left( \sum_{\mu=1}^n x_{\mu} \frac{\partial}{\partial x_{\mu}}
+ (D-1) \right) P_T^{[1]}(\xi_i^2)
\nonumber\\
&-&2 \delta_{\alpha \beta} \sum_{\stackrel{\scriptstyle j=1}{(j\neq i)}}^N 
\frac{m_j \xi_j^2} {\xi_i^2 - \xi_j^2}
\left( P_T^{[1]}(\xi_i^2)- P_T^{[1]}(\xi_j^2) \right)
- 2 \delta_{\alpha \beta} (m_i+1)  \xi_i^2 P_T^{[2]}(\xi_i^2), 
\end{eqnarray}
\noindent
and for (a) general type $(\varepsilon=0$)
\begin{eqnarray}
\mathcal{R}_{(\alpha,0), (\beta,0)}&=& 
\frac{1}{\sigma_n} \hat{\mathcal{R}}^{(0)}_{\alpha \beta} 
- \delta_{\alpha \beta} \left( \sum_{\mu=1}^n x_{\mu} \frac{\partial}{\partial x_{\mu}}
+ (D-1) \right) P^{[1]}_T(0) \nonumber \\
&+& 2 \delta_{\alpha \beta}  \sum_{j=1}^N m_j \left( P^{[1]}_T(0) - P_T^{[1]}(\xi_j^2) \right), 
\end{eqnarray}
\noindent
for (b) special type ($K=1$ and $\varepsilon=1$)
\begin{equation}
\mathcal{R}_{(1,0),(1,0)}
= - \sum_{\rho=1}^n \frac{1}{x_{\rho}} \frac{\partial P_T}{\partial x_{\rho}}
- 2 \sum_{i=1}^N m_i P_T^{[1]}(\xi_i^2),
\end{equation}
where $\hat{\mathcal{R}}^{(i)}_{\alpha \beta}$ and $\hat{\mathcal{R}}^{(0)}_{\alpha \beta}$
represent the Ricci components of $g^{(i)}$ and $g^{(0)}$ respectively.
Recall that $D=2n+2|m|+K$.


Since we are working in an orthonormal frame, 
the Einstein condition becomes $\mathcal{R}_{AB}=\Lambda \delta_{AB}$.
Thus, the K\"ahler metric $g^{(i)}$ (and the metric $g^{(0)}$ for general type)
must be Einstein, 
i.e. $\hat{\mathcal{R}}^{(i)}_{\alpha \beta}=\lambda^{(i)} \delta_{\alpha \beta}$,
($\hat{\mathcal{R}}^{(0)}_{\alpha \beta}= \lambda^{(0)} \delta_{\alpha \beta}$). 
We find the following result.\\

\noindent
{\bf{Theorem 3.}} Let $g^{(i)}~(i=1, \cdots, N)$ be  
$2m_i$-dimensional K\"ahler-Einstein metrics. Let $g^{(0)}$ be a $K$-dimensional 
Einstein metric if it is the general type.
Then the metric $g$ is Einstein if and only if $X_{\mu}$ takes the form
\begin{equation} \label{EC}
X_{\mu}(x_{\mu})
= x_{\mu}
\left( d_{\mu}
+ \int \mathcal{X}(x_{\mu}) \, x_{\mu}^{K-2} \prod_{i=1}^N( x_{\mu}^2 - \xi_i^2)^{m_i} \, d x_{\mu}
\right),
\end{equation}
where
\begin{equation}
\mathcal{X}(x) = \sum_{i=-\varepsilon}^n \alpha_i x^{2i}, \qquad
\alpha_n = - \Lambda,
\end{equation}
(a) general $(\varepsilon=0$), 
\begin{equation}
\alpha_0=(-1)^{n-1} \lambda^{(0)},
\end{equation}
(b) special ($K=1$ and $\varepsilon=1$), 
\begin{equation}
\alpha_0=(-1)^{n-1} 2c \sum_{j=1}^N \frac{m_j}{\xi_j^2}, \qquad
\alpha_{-1} = (-1)^{n-1} 2c.
\end{equation}
Here $\{ \alpha_k \}_{k=1,2,\dotsc, n-1}$ and $d_{\mu}$ are free parameters.
(When $K=0$, $\lambda^{(0)}$ is a free parameter.)
 
The Einstein constants $\lambda^{(i)}$ of $g^{(i)}$ are given by
\begin{equation}
\lambda^{(i)}=(-1)^{n-1} \mathcal{X}(\xi_i).
\end{equation}\\
\noindent
{\bf{Remark 3.}}
Note that $x_{\mu}^{\varepsilon} X_{\mu}(x_{\mu})$ is a polynomial. (For general type with $K=1$,
because any $1$-dimensional metric $g^{(0)}$
is flat, 
$\alpha_0=(-1)^{n-1} \lambda^{(0)} = 0$, thus there is no $\log x_{\mu}$ term.)
 
\noindent
{\bf{Remark 4.}} As in Remark 2, let us consider the $\xi_N \rightarrow 0$ limit.
For general type, it is easy to see that the polynomial $X_{\mu}$ \eqref{EC} 
is consistent with the limit.
For special type, with $c=\xi_N^2$,
$\alpha_0 \rightarrow (-1)^{n-1} 2 m_N$, $\alpha_{-1} \xi_N^{-2} \rightarrow (-1)^{n-1} 2$.
The Einstein constants of the K\"{a}hler-Einstein metric $g^{(N)}$ and the 
Einstein metric $g^{(0)}$ \eqref{Sasaki}
are  $\lambda^{(N)}=2(m_N+1)$ and $\lambda^{(0)} = 2m_N$ respectively.
The metric $g^{(0)}$ \eqref{Sasaki}, induced from the special type, 
is now a Sasaki-Einstein metric.\\
 

As a simple example, let us consider the special type metric with $n=1$ and $N=1$ for \eqref{EC}. 
Putting $x=i r$
together with the parameters 
$\{m_1=m, \xi_1=a, c=-a^2, d_1=(-1)^m 2M \}$
we have
\begin{equation}
P(r)= \left( -\frac{1}{r^2}+\frac{\Lambda}{2(m+1)} \right) (r^2+a^2)+\frac{2 M}{(r^2+a^2)^m}.
\end{equation}
The corresponding $2m+3$-dimensional metric is given by
\begin{equation}
g^{(2m+3)}=(r^2+a^2)g^{(2m)}-\frac{dr^2}{P(r)}+P(r)\theta_0^2+\frac{a^2}{r^2}
(\theta_0-r^2 \theta_1)^2.
\end{equation}
If we choose the Fubini-Study metric on $\mathbb{CP}^m$ for the Einstein-K\"ahler
metric $g^{(2m)}$, we reproduce the Kerr-de Sitter metric with mass $M$ and 
equal rotation parameter $a$ \cite{GLPP1}.


Finally, we briefly discuss the Einstein metrics over compact Riemannian manifolds
that are obtained from the metric \eqref{gCKY}. 
For general values of the parameters in \eqref{EC}
the metrics do not extend smoothly onto compact manifolds. 
For simplicity we consider an $n=1$ case of special type
\eqref{EC}. Let $(M^{(i)},g^{(i)},\omega^{(i)})~(i=1,\cdots,N)$ be 2$m_i$-dimensional
compact K\"ahler Einstein manifolds with positive first Chern class $c_1$.
One can write $c_1$ as $p_i \alpha_i$, where $\alpha_i$ is indivisible and $p_i$ is
a positive integer. Let $B=M^{(1)} \times \cdots \times M^{(N)}$ and $\pi_i$ be 
the projection map onto $M^{(i)}$. We will consider principal $T^2$ bundles over
$B$ which are classified by cohomology classes $\chi_a~(a=1,2)$ of the form
\begin{equation} \label{IC1}
\chi_a = \sum_{i=1}^N k_{i}^{(a)} \pi^{*}_{i} \alpha_i,~~~ k_{i}^{(a)} \in \mathbb{Z}.
\end{equation}

Now let us choose the roots $x_1$ and $x_2$ to the equation \eqref{EC}, $X(x)=0$.
We take the region $x_1 \le x \le x_2$ assuming that
$P(x)=X(x)/(x\prod_{i=1}^{N}(x^2-\xi_i^2)^{m_i}) \ge 0$. In order to avoid the singularity
at the boundaries $x=x_1$ and $x_2$ the following quantities must be integers:
\begin{equation} \label{IC2}
k^{(1)}_i=\frac{P'(x_1)}{1-(x_2/x_1)^2}\left(\xi_i-\frac{x_2^2}{\xi_i} \right)
\frac{p_i}{\lambda^{(i)}}, \qquad
k^{(2)}_i=\frac{P'(x_2)}{1-(x_1/x_2)^2}\left(\xi_i-\frac{x_1^2}{\xi_i} \right)
\frac{p_i}{\lambda^{(i)}}
\end{equation}
for $i=1,\cdots ,N$.
The integers $k_{i}^{(a)}$ may be identified with integral coefficients in \eqref{IC1}.

For example, we can obtain 5-dimensional Einstein metrics on $S^3$-bundle over $S^2$
constructed in \cite{HSY} as follows. Let us consider the case $B=\mathbb{CP}^1$.
For the real numbers $\nu_1$ and $\nu_2$ we put 
$\Lambda=4(1-\nu_1^2 \nu_2^2)/(2-\nu_1^2-\nu_2^2), c=\nu_1^2 \nu_2^2$ and $\xi_1=1$.
Then we have
\begin{equation}
P(x)
=\frac{(x^2-\nu_1^2)(x^2-\nu_2^2)(1-\Lambda x^2/4)}{x^2(x^2-1)}.
\end{equation}
The integral condition \eqref{IC2} is written as
\begin{equation} 
k^{(1)}=\frac{\nu_1(1-\nu_2^2)(2-\nu_2^2-\nu_1^2 \nu_2^2)}
{1+\nu_1^4 \nu_2^2+\nu_1^2 \nu_2^4-3 \nu_1^2 \nu_2^2}, \qquad
k^{(2)}=\frac{\nu_2(1-\nu_1^2)(2-\nu_1^2-\nu_1^2 \nu_2^2)}
{1+\nu_1^4 \nu_2^2+\nu_1^2 \nu_2^4-3 \nu_1^2 \nu_2^2},
\end{equation}
which is just the condition for the existence of Einstein metrics given in \cite{HSY}.

\vspace{5mm}

\noindent
{\bf{Acknowledgements}}

\vspace{3mm}

The work of YY is supported by the Grant-in Aid for Scientific
Research (No. 19540304 and No. 19540098)
from Japan Ministry of Education. 
The work of TO is supported by the Grant-in Aid for Scientific
Research (No. 19540304 and No. 20540278)
from Japan Ministry of Education.



\begin{thebibliography}{99}

\bibitem{CLP}
W. Chen, H. L\"{u} and C.N. Pope,
``General Kerr-NUT-AdS metrics in all dimensions,''
Class. Quant. Grav. \textbf{23} (2006) 5323-5340,
\texttt{arXiv:hep-th/0604125}.

\bibitem{HHOY}
N. Hamamoto, T. Houri, T. Oota and Y. Yasui,
``Kerr-NUT-de Sitter curvature in all dimensions,''
J. Phys. \textbf{A40} (2007) F177-F184,
\texttt{arXiv:hep-th/0611285}.

\bibitem{MP}
R.C. Myers and M.J. Perry,
``Black holes in higher dimensional space-times,"
Ann. Phys. {\bf{172}} (1986) 304-347.

\bibitem{HHT}
S.W. Hawking, C.J. Hunter and M.M. Taylor-Robinson,
``Rotation and the AdS/CFT correspondence,"
Phys. Rev. {\bf{D59}} (1999) 064005,
\texttt{arXiv:hep-th/9811056}.

\bibitem{GLPP1}
G.W. Gibbons, H. L\"{u}, D.N. Page and C.N. Pope,
``The general Kerr-de Sitter metrics in all dimensions,"
J. Geom. Phys. {\bf{53}} (2005) 49-73,
\texttt{arXiv:hep-th/0404008}.

\bibitem{GLPP2}
G.W. Gibbons, H. L\"{u}, D.N. Page and C.N. Pope,
``Rotating black holes in higher dimensions with a cosmological constant,"
Phys. Rev. Lett. {\bf{93}} (2004) 171102,
\texttt{arXiv:hep-th/0409155}.

\bibitem{CLP1}
W. Chen, H. L\"{u} and C.N. Pope,
``Kerr-de Sitter Black Holes with NUT Charges,"
Nucl. Phys. \textbf{B762} (2007) 38-54,
\texttt{arXiv:hep-th/0601002}.


\bibitem{FK}
V.P. Frolov and D. Kubiz\v{n}\'{a}k,
```Hidden' Symmetries of Higher Dimensional Rotating Black Holes,''
Phys. Rev. Lett. \textbf{98} (2007) 11101,
\texttt{arXiv:gr-qc/0605058}.

\bibitem{KF}
D. Kubiz\v{n}\'{a}k and V.P. Frolov,
``Hidden Symmetry of Higher Dimensional Kerr-NUT-AdS Spacetimes,''
Class. Quant. Grav. \textbf{24} (2007) F1-F6,
\texttt{arXiv:gr-qc/0610144}.

\bibitem{PKVK}
D.N. Page, D. Kubiz\v{n}\'{a}k, M. Vasudevan and P. Krtou\v{s},
``Complete Integrability of Geodesic Motion in General Kerr-NUT-AdS Spacetimes,''
Phys. Rev. Lett. \textbf{98} (2007) 061102,
\texttt{arXiv:hep-th/0611083}.

\bibitem{FKK}
V.P. Frolov, P. Krtou\v{s} and D. Kubiz\v{n}\'{a}k,
``Separability of Hamilton-Jacobi and Klein-Gordon Equations
in General Kerr-NUT-AdS Spacetimes,''
JHEP \textbf{0702} (2007) 005,
\texttt{arXiv:hep-th/0611245}.

\bibitem{KKPF}
P. Krtou\v{s}, D. Kubiz\v{n}\'{a}k, D.N. Page and V.P. Frolov,
``Killing-Yano Tensors, Rank-2 Killing Tensors,
and Conserved Quantities in Higher Dimensions,''
JHEP \textbf{0702} (2007) 004,
\texttt{arXiv:hep-th/0612029}.



\bibitem{KKPV}
P. Krtou\v{s}, D. Kubiz\v{n}\'{a}k, D.N. Page and M. Vasudevan,
``Constants of Geodesic Motion in Higher-Dimensional Black-Hole Spacetime,''
Phys. Rev. \textbf{D76} (2007) 084034,
\texttt{arXiv:hep-th/0707.0001}.

\bibitem{HOY}
T. Houri, T. Oota and Y. Yasui,
``Closed conformal Killing-Yano tensor and geodesic integrability,"
J. Phys. A: Math. Theor. \textbf{41} (2008) 025204,
\texttt{arXiv:hep-th/0707.4039}.

\bibitem{HOY3}
T. Houri, T. Oota and Y. Yasui,
``Closed conformal Killing-Yano tensor and Kerr-NUT-de Sitter uniqueness,"
Phys. Lett. \textbf{B656} (2007) 214-216,
\texttt{arXiv:0708.1368[hep-th]}.

\bibitem{OY2}
T. Oota and Y. Yasui,
``Separability of Dirac equation in higher dimensional Kerr-NUT-de Sitter
spacetime,"
Phys. Lett. \textbf{B659} (2008) 688-693,
\texttt{arXiv:0711.0078[hep-th]}.

\bibitem{KF07}
D. Kubiz\v{n}ak and V.P. Frolov,
``Stationary strings and branes in the higher-dimensional Kerr-NUT-(A)dS
spacetimes,"
JHEP \textbf{0802} (2008) 007,
\texttt{arXiv:0711.2300[hep-th]}.

\bibitem{SK}
A. Sergyeyev and P. Krtou\v{s},
``Complete Set of Commuting Symmetry Operators for Klein-Gordon Equation 
in Generalized Higher-Dimensional Kerr-NUT-(A)dS Spacetimes,"
Phys. Rev. \textbf{D77} (2008) 044033,
\texttt{arXiv:0711.4623[hep-th]}.

\bibitem{flor08}
V.P. Frolov,
``Hidden Symmetries of Higher-Dimensional Black Hole Spacetimes,"
\texttt{arXiv:0712.4157[gr-qc]}.


\bibitem{FK08}
V.P. Frolov and D. Kubiz\v{n}\'{a}k,
``Higher-Dimensional Black Holes: Hidden Symmetries and Separation of Variables,"
\texttt{arXiv:0802.0322[hep-th]}.

\bibitem{CFK}
P. Connell, V.P. Frolov and D. Kubiz\v{n}\'{a}k,
``Solving parallel transport equations in the higher-dimensional Kerr-NUT-(A)dS
spacetimes,"
\texttt{arXiv:0803.3259[gr-qc]}.

\bibitem{KFK08}
P. Krtou\v{s}, V.P. Frolov and D. Kubiz\v{n}\'{a}k,
``Hidden Symmetries of Higher Dimensional Black Holes
and Uniqueness of the Kerr-NUT-(A)dS spacetime,''\\
\texttt{arXiv:0804.4705[hep-th]}.


\bibitem{tac}
S. Tachibana,
``On conformal Killing tensor in a Riemannian space,''
T\^{o}hoku Math. J. \textbf{21} (1969) 56-64.

\bibitem{HOY2}
T. Houri, T. Oota and Y. Yasui,
``Closed conformal Killing-Yano tensor and uniqueness of generalized Kerr-NUT-de Sitter spacetime,"
\texttt{arXiv:0805.3877[hep-th]}.

\bibitem{CLPP}
M. Cveti\v{c}, H. L\"{u}, D.N. Page and C.N. Pope,
Phys. Rev. Lett. \textbf{95} (2005) 071101,
``New Einstein-Sasaki Spaces in Five and Higher Dimensions,"
\texttt{arXiv:hep-th/0504225}.

\bibitem{CLPP2}
M. Cveti\v{c}, H. L\"{u}, D.N. Page and C.N. Pope,
``New Einstein-Sasaki and Einstein Spaces from Kerr-de Sitter,"
\texttt{arXiv:hep-th/0505223}.

\bibitem{HSY}
Y. Hashimoto, M. Sakaguchi and Y. Yasui,
``New infinite series of Einstein metrics on sphere bundles from AdS black holes,"
Commun. Math. Phys. \textbf{257} (2005) 273-285,
\texttt{arXiv:hep-th/0402199}.












\end{thebibliography}
\end{document}